\documentclass[a4paper,11pt]{article}
\usepackage{graphicx,epsfig,picins}
\usepackage{fancyhdr,fancybox}
\usepackage{indentfirst}
\usepackage{titlesec}
\usepackage{verbatim}
\usepackage[sort&compress,numbers]{natbib}
\usepackage{array,dcolumn,tabularx}
\usepackage{setspace}
\usepackage{geometry}
\usepackage{extarrows,chemarrow,xypic} 
\usepackage[small]{caption2}

\usepackage{times}     

\usepackage{latexsym}
\usepackage{amsmath}                 
\usepackage{amssymb}                 
\usepackage{amsbsy}
\usepackage{amsthm}
\usepackage{amsfonts}
\usepackage{mathrsfs}                
\usepackage{bm}                      
\usepackage{relsize}                 
\usepackage{hyperref}  

\newcommand{\paperfont}{\fontsize{11pt}{1.1\baselineskip}\selectfont}
\begin{document}

\theoremstyle{definition}
\makeatletter
\thm@headfont{\bf}
\makeatother
\newtheorem{result}{Basic Result}

\lhead{}
\rhead{}
\lfoot{}
\rfoot{}

\renewcommand{\refname}{References}
\renewcommand{\figurename}{Figure}
\renewcommand{\tablename}{Table}
\renewcommand{\proofname}{Proof}

\title{\textbf{Modeling stochastic phenotype switching and bet-hedging in bacteria: stochastic nonlinear dynamics and critical state identification}}
\author{Chen Jia$^{1,2}$,\;\;\;Min-Ping Qian$^1$,\;\;\;Yu Kang$^3$,\;\;\;Da-Quan Jiang$^{1,4,*}$ \\
\footnotesize $^1$LMAM, School of Mathematical Sciences, Peking University, Beijing 100871, P.R. China \\
\footnotesize $^2$Beijing International Center for Mathematical Research, Beijing 100871, P.R. China \\
\footnotesize $^3$CAS Key Laboratory of Genome Sciences and Information, Beijing Institute of Genomics, \\
\footnotesize Chinese Academy of Sciences, Beijing 100101, P.R. China; \\
\footnotesize $^4$Center for Statistical Science, Peking University, Beijing 100871, P.R. China\\
\footnotesize $^*$Correspondence: jiangdq@math.pku.edu.cn}
\date{}                              
\maketitle                           

\paperfont

\begin{abstract}
Fluctuating environments pose tremendous challenges to bacterial populations. It is observed in numerous bacterial species that individual cells can stochastically switch among multiple phenotypes for the population to survive in rapidly changing environments. This kind of phenotypic heterogeneity with stochastic phenotype switching is generally understood to be an adaptive bet-hedging strategy. Mathematical models are essential to gain a deeper insight into the principle behind bet-hedging and the pattern behind experimental data. Traditional deterministic models cannot provide a correct description of stochastic phenotype switching and bet-hedging, and traditional Markov chain models at the cellular level fail to explain their underlying molecular mechanisms. In this paper, we propose a nonlinear stochastic model of multistable bacterial systems at the molecular level. It turns out that our model not only provides a clear description of stochastic phenotype switching and bet-hedging within isogenic bacterial populations, but also provides a deeper insight into the analysis of multidimensional experimental data. Moreover, we use some deep mathematical theories to show that our stochastic model and traditional Markov chain models are essentially consistent and reflect the dynamic behavior of the bacterial system at two different time scales. In addition, we provide a quantitative characterization of the critical state of multistable bacterial systems and develop an effective data-driven method to identify the critical state without resorting to specific mathematical models. \\

\end{abstract}

\section*{Introduction}
Bacteria in the wild exist in ever-changing environments and have to surmount the challenges posed by environmental fluctuations. Numerous experiments have confirmed that multiple distinct phenotypes can coexist within an isogenic bacterial population \cite{kussell2005phenotypic, smits2006phenotypic, dubnau2006bistability, avery2006microbial, dhar2007microbial, lu2007phenotypic, veening2008bistability, fraser2009chance, jablonka2009transgenerational, snijder2011origins}. This phenotypic heterogeneity in genetically identical cells has received increasing attention in recent years since it could help the bacteria to survive in rapidly changing environments \cite{mao2014slow, rulands2014specialization}. In the framework of traditional population genetics, a bacterial population enhances its fitness via genetic changes caused by mutation or recombination. However, extracellular conditions can change so rapidly that adaptation only by mutation or recombination would be too slow. One solution to this problem is to allow individual cells to stochastically switch among multiple phenotypes without genetic changes, a phenomenon widely known as stochastic phenotype switching \cite{acar2008stochastic, salathe2009evolution, leisner2008stochastic, gaal2010exact, libby2011exclusion, rainey2011evolutionary}. Generally, the multiple phenotypes within an isogenic bacterial population result from the multiple steady-state expression levels of a group of stress-related genes. Such kind of gene expression pattern with multiple steady-state expression levels are widely known as multistability \cite{dubnau2006bistability, veening2008bistability}.

Phenotypic heterogeneity is a widespread phenomenon in the bacterial realm. Examples of phenotypic heterogeneity include lactose utilization in \emph{Escherichia coli} \cite{ozbudak2004multistability}, competence development in \emph{Bacillus subtilis} \cite{suel2006excitable, tsang2006exciting, schultz2007molecular}, sporulation in \emph{Bacillus subtilis} \cite{sonenshein2002bacillus, errington2003regulation, morohashi2007model, de2010heterochronic}, and persistence in \emph{Mycobacterium tuberculosis} \cite{sureka2008positive, gefen2009importance, ghosh2011phenotypic}. The potential function of phenotypic heterogeneity with stochastic phenotype switching is generally understood to be a bet-hedging strategy \cite{veening2008bistability, veening2008bet, beaumont2009experimental, rulands2014specialization}, a term originating from finance. In response to fluctuating environments, a heterogeneous bacterial population could optimize its fitness by altering the proportion of cells in each subpopulation via stochastic phenotype switching to achieve an optimal `investment portfolio'.

To study the evolution of heterogeneous bacterial populations, a number of Markov chain models have been proposed at the cellular level \cite{kussell2005phenotypic, wolf2005diversity, lu2007phenotypic, salathe2009evolution, gaal2010exact, gupta2011stochastic, libby2011exclusion, zhou2013population}. These models assumed \emph{a priori} that the bacterial population has multiple distinct phenotypes. In these models, each phenotype is modeled as a state of the Markov chain and stochastic phenotypic switching is modeled as the state transition of the Markov chain. However, these models take phenotypic heterogeneity and stochastic phenotypic switching for granted and fail to account for their underlying molecular mechanisms.

Recent research has demonstrated that phenotypic heterogeneity within isogenic bacterial populations often results from the feedback circuitry of the gene regulatory network \cite{karmakar2007positive, mitrophanov2008positive}. To account for the molecular mechanism of phenotypic heterogeneity, a number of deterministic models have been proposed at the molecular level \cite{ozbudak2004multistability, mantzaris2007single, schultz2007molecular, sureka2008positive, ghosh2011phenotypic}. In these models, different steady states of gene expression are described as different stable fixed points (attractors) of a deterministic system composed of several ordinary differential equations which are written down based on the regulatory relationship of the gene network. However, deterministic models cannot provide a correct description of many important experimental phenomena, such as stochastic phenotype switching and bet-hedging. In every deterministic model, if the expression level of an individual cell lies in an attraction basin at a particular time, it will never leave this attraction basin and thus phenotype switching will never occur.

Although deterministic models can give rise to multiple attractors and attraction basins, they do not allow transitions among different attraction basins. One solution to this problem is to consider stochastic effects, which allow the system to transition among different attraction basins and thus drive stochastic phenotype switching. This fact is analogous to the simulated annealing techniques in optimization problems, in which noise is indispensable to make the search escape from the trap of local minimum points and reach the global minimum point. In order to better understand the role that stochastic effects play in bistable systems, Qian and coworkers studied the relations between deterministic and stochastic nonlinear dynamics in great detail \cite{vellela2009stochastic, qian2010chemical, qian2011nonlinear, ge2011non, qian2012cooperativity, qian2012mesoscopic}. However, their models are usually so abstract and oversimplified that they cannot be directly applied to practical problems with experimental data and observations.

Stochastic effects are extremely important not simply because they are indispensable for the model to generate phenotype switching, but because gene expression is an inherently stochastic process. Recent developments of single-cell and single-molecule experiments have shown that many important cellular processes, such as transcription, translation, replication, and gene regulation, are inherently stochastic \cite{mcadams1997stochastic, elowitz2002stochastic, ozbudak2002regulation, paulsson2004summing, kaern2005stochasticity, raser2005noise, cai2006stochastic, yu2006probing, xie2008single, sanchez2013regulation}. Due to stochastic effects, the expression levels of the stress-related genes in a multistable system will have a multimodal distribution.

In this paper, we propose a unified nonlinear stochastic model of multistable bacterial systems at the molecular level based on a core double-positive-feedback gene network. By studying its stochastic nonlinear dynamics, we show that our model not only provides a clear description of phenotypic heterogeneity, stochastic phenotype switching, and bet-hedging within isogenic bacterial populations, but also provides a deeper insight into the analysis of multidimensional experimental data, such as gene expression data and the data of more comprehensive indicators like the forward scatter (FSC) and the side scatter (SSC) measured by flow cytometry.

Next, we use the mathematical tool of large deviation theory established by Freidlin and Wentzell \cite{freidlin2012random} to show that every multistable dynamical system under a small random perturbation can be approximated by a Markov chain with multiple states, each corresponding to an attraction basin of the multistable system. In this way, our stochastic model at the molecular level can be reduced to a Markov chain model at the cellular level. This justifies the wide applications of previous Markov chain models of population evolution, in particular, the Markov chain model proposed by Lander and coworkers \cite{gupta2011stochastic} about the dynamics of the phenotypic proportions in human breast cancer cell lines.

In addition, we point out a widespread misunderstanding on the analysis of gene expression data, inspired by our recent work about antibiotic resistance in \emph{Escherichia coli}. Previous studies tended to think that phenotypic heterogeneity can be identified by the multistable expression of a single pivotal gene (reviewed in \cite{smits2006phenotypic, veening2008bistability}). However, phenotypic heterogeneity in bacterial populations often results from the interaction of a group of stress-related genes. We use simulation results to show that in many cases, the expression data of a group of genes give rise to an apparent multimodal distribution, however, we cannot observe the multistable expression if we only focus on the expression data of a single gene. This suggests that the traditional method to identify phenotypic heterogeneity by measuring the expression of a single pivotal gene is sometimes ineffective.

Finally, we use our stochastic model to provide an answer to the important question of identifying the critical state of multistable bacterial systems. In our stochastic model, there is a saddle lying on the boundary of two adjacent attraction basins which characterizes a critical state between two steady states of gene expression. The critical state is not targeted in the previous work since it is rarely observed in experiments and cannot be estimated by simple statistical analysis of gene expression data. However, the identification of the critical state has drawn increasing attention in recent years since it is closely related to the early diagnosis of complex diseases \cite{kim2007hidden, kellershohn2001prion}. In this paper, we develop an effective method to identify the critical state of multistable bacterial systems using the time-course data of gene expression without resorting to specific mathematical models.

\section*{Model}
\begin{figure}[!htb]
\begin{center}
\centerline{\includegraphics[width=0.8\textwidth]{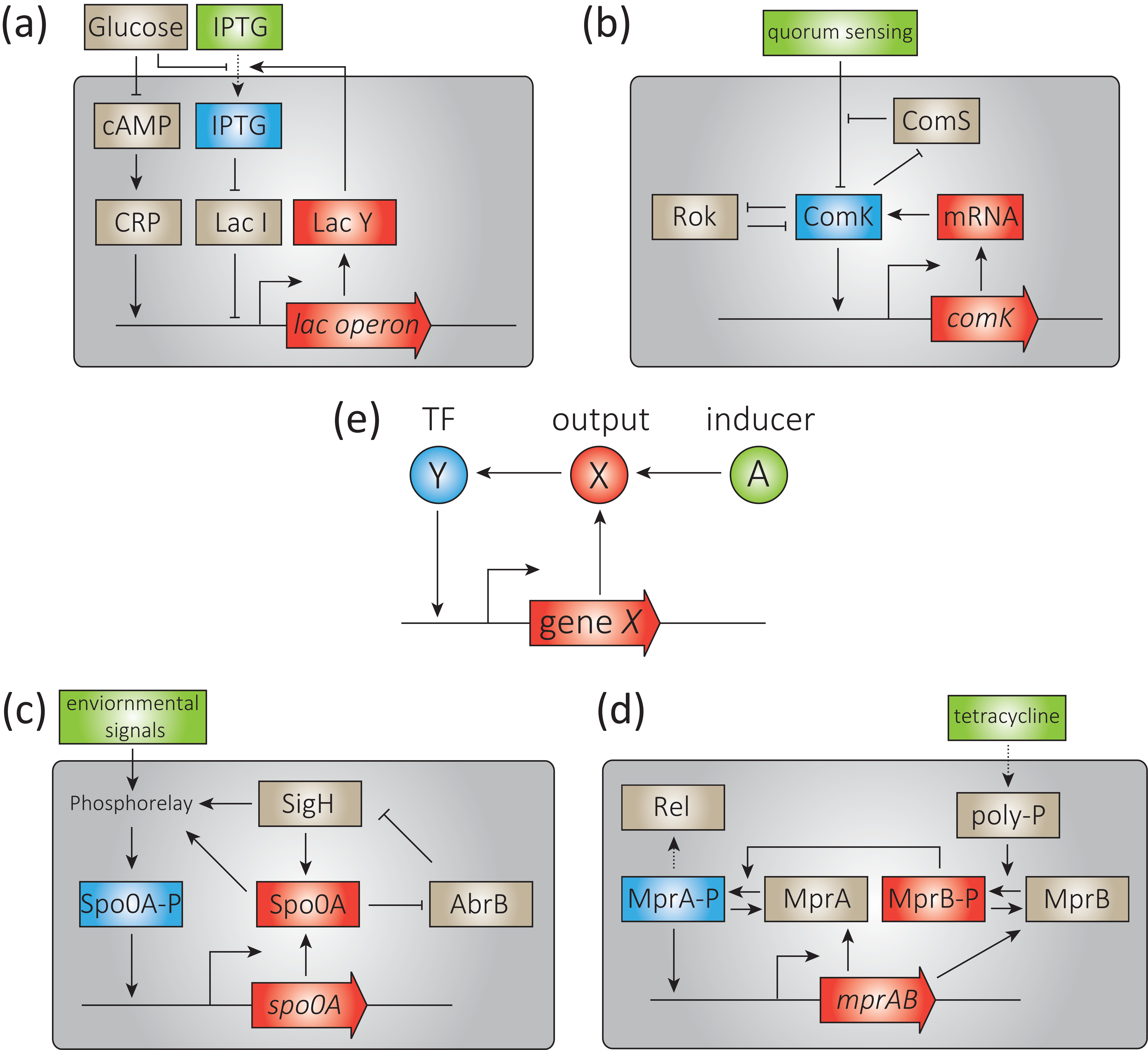}}
\caption{\textbf{Schematic models of bacterial systems with phenotypic heterogeneity.}
\textbf{a-d.} Examples of naturally occurring bacterial systems with phenotypic heterogeneity.
\textbf{a.} Lactose utilization in \emph{E. coli} \cite{ozbudak2004multistability}.
\textbf{b.} Competence development in \emph{B. subtilis} \cite{suel2006excitable}.
\textbf{c.} Sporulation in \emph{B. subtilis} \cite{sonenshein2002bacillus}.
\textbf{d.} Persistence in \emph{M. tuberculosis} \cite{ghosh2011phenotypic}.
\textbf{e.} The core double-positive-feedback gene network shared by a-d. X (\emph{red}) is the product (protein or mRNA) of a pivotal stress-related gene. Y (\emph{blue}) is a transcription factor which activates the expression of gene \emph{X}. A (\emph{green}) is an inducer whose concentration reflects extracellular environmental conditions.}\label{network}
\end{center}
\end{figure}

In natural bacterial systems, phenotypic heterogeneity and stochastic phenotype switching always originate from the feedback circuitry of the regulatory network which governs a group of stress-related genes. In order to better understand the general principles behind phenotypic heterogeneity, we illustrate the gene regulatory networks that govern some best-understood multistable systems in bacteria (Fig. \ref{network}(a)-(d)). A crucial similarity shared by these examples is that the wiring of the gene regulatory network forms a double-positive-feedback loop. In order to establish a unified model of these bacterial systems, we focus on the core double-positive-feedback gene network depicted in Fig. \ref{network}(e), where, protein X is the product of a pivotal stress-related gene, i.e., gene $X$, protein Y is a transcription factor which activates the expression of gene $X$, and $A$ is an inducer whose concentration reflects the fluctuations in extracellular environmental conditions, such as fluctuations in temperature, pH, and concentrations of nutrients and toxins \cite{sonenshein2002bacillus, sureka2008positive}.

We use lowercase letters $x$, $y$, and $a$ to denote the concentrations of X, Y, and A, respectively. Since gene expression is an inherently stochastic process, the dynamics of $x$ and $y$ can be described by the following two-dimensional system of stochastic differential equations (SDEs):
\begin{equation}\label{2d}\left\{
\begin{split}
\dot{x} &= -\alpha(x - F(a,y)) + \sqrt{2\epsilon}\xi_x, \\
\dot{y} &= -\beta(y - G(x)) + \sqrt{2\eta}\xi_y,
\end{split}\right.
\end{equation}
where $F(a,y)$ describes the activation of protein X by inducer A and protein Y, $G(x)$ describes the activation of protein Y by protein X, and $\alpha$ and $\beta$ are two parameters characterizing the response speeds of proteins X and Y, respectively. In addition, $\xi_x$ and $\xi_y$ are two independent standard white noises satisfying $\langle\xi_x(t)\rangle = \langle\xi_y(t)\rangle = 0$ and $\langle\xi_x(t)\xi_x(t')\rangle = \langle\xi_y(t)\xi_y(t')\rangle = \delta(t'-t)$. Since the fluctuations in the levels of proteins X and Y can be different, we use two noise levels $\epsilon$ and $\eta$ to describe their stochastic fluctuations. We emphasize here that noise in gene regulatory networks generally comes from a great number of sources and may not be subsumed into white noises. Fortunately, the specific noise distributions will hardly affect the main results of this paper. To make our discussion friendly to both theoretical and experimental biologists, we would like to use white noises to describe noise in gene expression. For the rationality of this assumption, please see \emph{Discussion}.

If we ignore stochastic effects, then the stochastic system \eqref{2d} can be reduced to the following deterministic system as the two noise levels, $\epsilon$ and $\eta$, tend to zero:
\begin{equation}\label{ODE}
\dot{x} = -\alpha(x - F(a,y)),~~~\dot{y} = -\beta(y - G(x)).
\end{equation}
We note that the fixed points of this deterministic system are the solutions to the following equations: $x - F(a,G(x)) = 0$ and $y=G(x)$. In natural bacterial systems, the most common expression of $F(a,G(x))$ has the following form (see \emph{Supplementary Information}):
\begin{equation}
F(a,G(x)) = \frac{\gamma x^n}{K+x^n} +(\mu a+\delta),
\end{equation}
where the Hill function $\gamma x^n/(K+x^n)$ with $n>1$ describes the activation of gene $X$ by protein Y, the term $\mu a$ describes the activation of protein X by inducer A, and the term $\delta$ describes a basal expression level of gene $X$ independent of the activation of protein Y.

It turns out that the deterministic system \eqref{ODE} has one or three fixed points under different inducer concentrations (Fig. \ref{SDEmodel}(a)). To be specific, the inducer concentration $a$ has two threshold levels, $a_0$ and $a_1$. If $a<a_0$ or $a>a_1$, the system has only one fixed point (Fig. \ref{SDEmodel}(a)). If $a<a_0$, the only attractor $(x_L,y_L)$ describes the phenotype of low-expressing cells in which gene \emph{X} is inactivated. If $a>a_1$, the only attractor $(x_H,y_H)$ describes the phenotype of high-expressing cells in which gene \emph{X} is activated. If $a_0<a<a_1$, however, the system has three fixed points, including two attractors, $(x_L,y_L)$ and $(x_H,y_H)$, and a saddle $(x_M,y_M)$ (Fig. \ref{SDEmodel}(a),(b)), where the two attractors describe the phenotypes of low- and high-expressing cells, respectively, whereas the saddle is a critical state between the two steady states of gene expression. Mathematically, each attractor of a deterministic system has an attraction basin, and two adjacent attraction basins are separated by a boundary (Fig. \ref{SDEmodel}(b)).

\section*{Results}

\subsection*{Phenotypic heterogeneity and bet-hedging}
In the recent decade, single-cell and single-molecule experiments have made significant progresses and shown that gene expression is an inherently stochastic process. Although deterministic models, such as the deterministic system \eqref{ODE}, can give rise to multiple attractors and attraction basins, they cannot provide a correct description of stochastic phenotype switching and bet-hedging. These two facts show that the reduction from the stochastic system \eqref{2d} to the deterministic system \eqref{ODE} is inappropriate.

In the following discussion, we focus on the stochastic system \eqref{2d} in which the two noise levels, $\epsilon$ and $\eta$, are strictly positive. Although the system does not satisfy detailed balance, we can still obtain an approximate steady-state probability distribution $p_s(a,x,y)$ of the system (see \emph{Supplementary Information}), which has the following form:
\begin{equation}
p_s(a,x,y) = \frac{1}{Z}\exp\left\{-\frac{1}{\epsilon}U(a,x,y)\right\},
\end{equation}
where $Z$ is a normalization constant and $U(a,x,y)$ is an approximate global potential, also called landscape, of the system defined as
\begin{equation}\label{potential}
U(a,x,y) = \frac{\epsilon\beta}{2\eta}(y-G(x))^2+\alpha\int_0^x(u-F(a,G(u))du.
\end{equation}

\begin{figure}[!htb]
\begin{center}
\centerline{\includegraphics[width=0.8\textwidth]{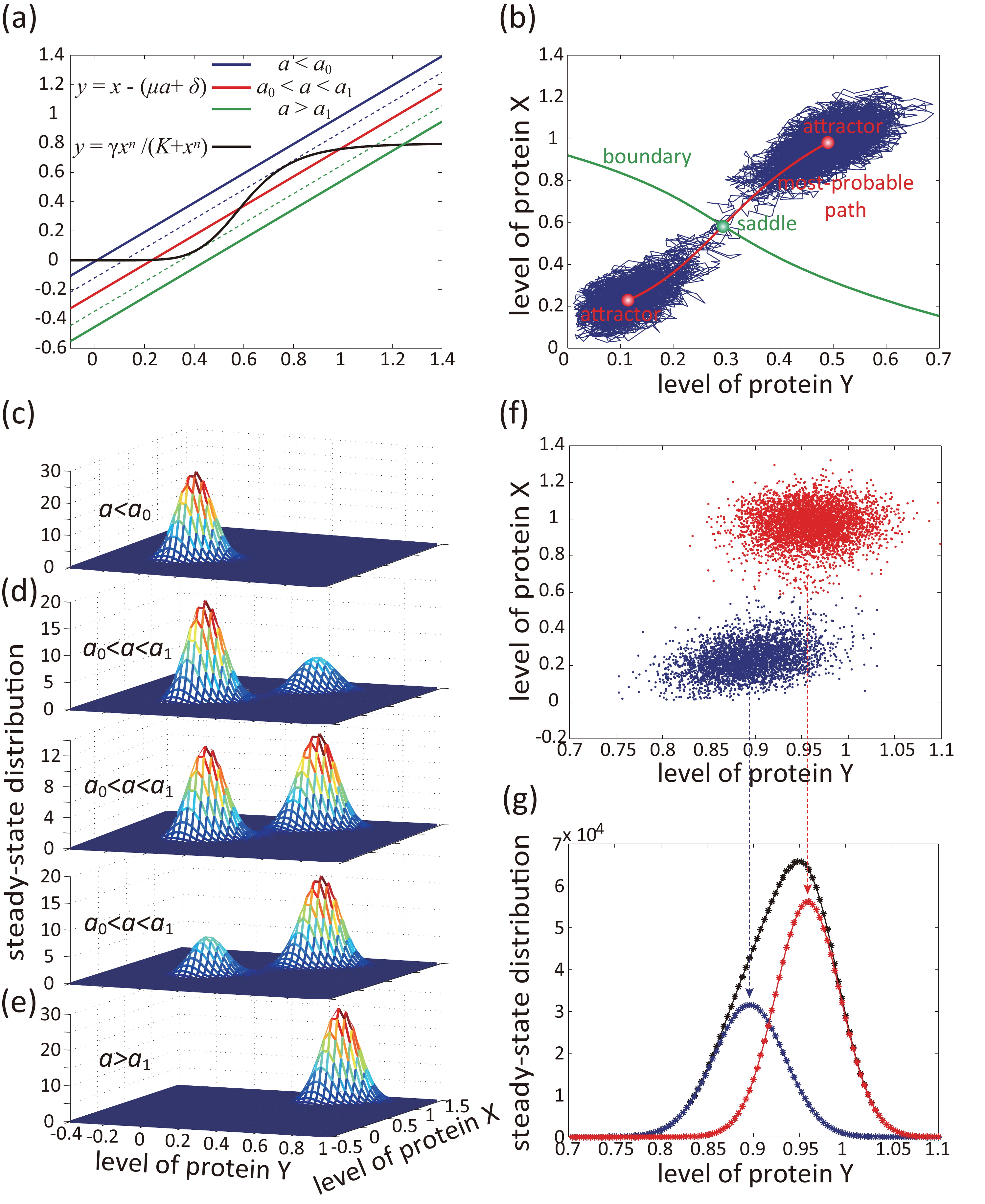}}
\caption{\textbf{Stochastic description of multistability.}
\textbf{a.} Numbers of fixed points under different inducer concentrations. The curve represents the Hill function $y = \gamma x^n/(K+x^n)$ and the lines represent the function $y = x-(\mu a+\delta)$. The intersections of the curve and the line give the positions of the fixed points.
\textbf{b.} The simulation data of the time course of the levels of proteins X and Y in a single cell. When $a_0<a<a_1$, the deterministic model has three fixed points, including two attractors and a saddle, which lies on the boundary of two attraction basins. The expression levels of the cell stay around the two attractors at most times and cross the boundary around the saddle.
\textbf{c-e.} The steady-state probability distribution of the levels of proteins X and Y.
\textbf{c.} When $a<a_0$, the steady-state gene expression levels have a monomodal distribution, which attains its unique maximum at the attractor $(x_L,y_L)$.
\textbf{d.} When $a_0<a<a_1$, the steady-state gene expression levels have a bimodal distribution, which attains two maxima at both the two attractors, $(x_L,y_L)$ and $(x_H,y_H)$.
\textbf{e.} When $a>a_1$, the steady-state gene expression levels have a monomodal distribution, which attains its unique maximum at the attractor $(x_H,y_H)$.
\textbf{f.} The simulation data of the steady-state levels of proteins X and Y in 500,000 virtual cells under a set of model parameters. The two-dimensional gene expression data are distributed around two attractors and thus lead to an apparent bimodal distribution.
\textbf{g.} The marginal distribution of the level of protein Y based on the two-dimensional gene expression data. The blue and red curves represent the marginal distributions of low- and high-expressing cells, respectively, and the black curve represents the marginal distribution of all cells. There is an obvious overlap between the marginal distributions of low- and high-expressing cells, resulting in a monomodal overall marginal distribution.
}\label{SDEmodel}
\end{center}
\end{figure}

We make a crucial observation that the fixed points of the deterministic system \eqref{ODE} are exactly the solutions to the equation $\partial_xU(a,x,y) = \partial_yU(a,x,y) = 0$. This shows that the attractors of the deterministic system are the local minimum points of the approximate global potential $U(a,x,y)$ and thus are the local maximum points of the steady-state distribution $p_s(a,x,y)$. From Fig. \ref{SDEmodel}(c)-(e), we see that the steady-state distribution of the levels of proteins X and Y is controlled by the inducer concentration $a$. If $a<a_0$, the steady-state distribution has only the left peak, suggesting that the bacterial population contains almost exclusively low-expressing cells under favorable conditions (Fig. \ref{SDEmodel}(c)). If $a_0<a<a_1$, the steady-state distribution has both the left and right peaks, each corresponding to a phenotype. With the increase of the inducer concentration, a larger fraction of cells will switch from the low- to the high-expressing subpopulation to maximize survival (Fig. \ref{SDEmodel}(d)). If $a>a_1$, the left peak of the steady-state distribution disappears, suggesting that the bacterial population contains almost exclusively high-expressing cells under unfavorable conditions (Fig. \ref{SDEmodel}(e)). The above discussion clearly explains how the bet-hedging strategy could help the bacterial population better adapt to rapidly changing environmental conditions.

Generally, the steady-state gene expression levels in a multistable bacterial system have a multimodal distribution, which can be viewed as the superposition of multiple monomodal distributions, each concentrated within an attraction basin. The attractors are the locally most-probable states and thus are most likely to be observed in experiments. Based on the stochastic system \eqref{2d}, we simulate the time course of the levels of proteins X and Y in a single cell (Fig. \ref{SDEmodel} (b)). The simulation result shows that the gene expression data are generally distributed around the attractors and are rarely distributed around the boundary of the attraction basins. These facts clearly show that each phenotype of a bacterial population cannot be simply described as an attractor of the deterministic model, but should be understood as a monomodal distribution concentrated within an attraction basin.

\subsection*{A widespread misunderstanding on the analysis of gene expression data}
Phenotypic heterogeneity in isogenic bacterial populations often results from the interaction of a group of stress-related genes and biochemical species. Previous studies tended to think that phenotypic heterogeneity can be identified by the multistable expression of a single pivotal gene (reviewed in \cite{smits2006phenotypic, veening2008bistability}). In experiments, however, it often happens that the steady-state expression data of a single gene does not display a multimodal distribution, and only when a subpopulation of cells are sorted out to start from some extreme initial conditions, the multimodal distribution can be observed at certain times before reaching the steady state. Thus it is rather difficult to determine whether the bacterial population has multiple phenotypes or not. To explain these experimental phenomena, we point out that phenotypic heterogeneity often results from the interaction of a group of stress-related genes and may not be observed if we only focus on the expression data of a single gene. In our recent study about antibiotic resistance in \emph{Escherichia coli} (unpublished work), we found that the expression data of the hydrolase gene only lead to a monomodal distribution, whereas the expression data of a group of stress-related genes lead to an apparent multimodal distribution.

We now use our stochastic model to account for this interesting phenomenon. Based on the stochastic system \eqref{2d}, we simulate the steady-state levels of proteins X and Y in 500,000 virtual cells under a set of model parameters (Fig. \ref{SDEmodel}(f)). From the simulation result, we see that the two-dimensional gene expression data are distributed around two attractors in the phase plane and lead to a bimodal distribution, which can be viewed as the superposition of two monomodal distributions. Although these two monomodal distributions are concentrated within two different attraction basins in the phase plane, there is an obvious overlap between their marginal distributions, whose superposition, which represents the steady-state distribution of the level of protein Y, has only one peak (Fig. \ref{SDEmodel}(g)). This suggests that the traditional idea to identify phenotypic heterogeneity by the multistable expression of a single pivotal gene is sometimes ineffective. The above discussion also shows that our stochastic model can help us gain a deeper insight into the pattern behind multidimensional experimental data.

\subsection*{From the molecular level to the cellular level}
We have seen that our stochastic model provides a clear description of phenotypic heterogeneity and bet-hedging within isogenic bacterial populations at the molecular level. However, more widely used models in the previous work are Markov chain models at the cellular level. These models assume \emph{a priori} that the bacterial population has multiple distinct phenotypes, each of which corresponds to a state of the Markov chain and can switch to other phenotypes with certain transition rates. This raises the question of whether the two kinds of models, our stochastic model at the molecular level and the Markov chain models at the cellular level, are consistent in some way or not. We now use the mathematical tool of large deviation theory established by Freidlin and Wentzell \cite{freidlin2012random} to answer this question.

The Freidlin-Wentzell theory is mainly concerned about the dynamic behavior of a multistable dynamical system under random perturbations, such as the stochastic system \eqref{2d}, when the noise level is not too large. The conclusions of the Freidlin-Wentzell theory are not very intuitive at first sight and the proofs of them are rather tedious. Readers who are interested in the mathematical aspects of the Freidlin-Wentzell theory may refer to \cite{freidlin2012random, olivieri2005large}. To make readers understand this useful mathematical tool, we would like to list the major results of the Freidlin-Wentzell theory as follows.

\begin{result}
No matter how small the noise level is, the accumulation of the stochastic forces will make the system escape from the trap of an attraction basin and enter another attraction basin. Before the system escapes from an attraction basin, it will spend most of the time staying around the attractor and spend little time staying around the boundary of the attraction basin. These facts can be seen from our numerical simulation in Fig \ref{SDEmodel}(b) and Fig. \ref{switch}(a).
\end{result}

\begin{result}\label{where}
Each point $x$ in an attraction basin has a local potential $V(x)$ called the quasi-potential. If the system has a global potential $U(x)$, as in Equation \eqref{potential}, then the quasi-potential $V(x)$ can be calculated explicitly as
\begin{equation}
V(x) = 2(U(x)-U(x_0)),
\end{equation}
where $x_0$ is the attractor in this attraction basin. When the system escapes from one attraction basin to another, it must cross the boundary around a specific point $y_0$ where the quasi-potential attains its minimum. In most cases, the minimum point $y_0$ of the quasi-potential on the boundary is exactly the saddle of the system. These facts can be seen from our numerical simulation in Fig. \ref{SDEmodel}(b).
\end{result}

\begin{result}\label{when}
The time needed for the system to escape from an attraction basin is referred to as the escape time. The escape time $T$ from an attraction basin approximately follows an exponential distribution. The mean escape time $\langle T\rangle$, which is approximately the time constant of the exponential distribution, has the form of
\begin{equation}
\langle T\rangle\doteq k\exp\left(\frac{1}{2\epsilon}V_0\right),
\end{equation}
where $k$ is a positive constant, $\epsilon$ is the noise level, and $V_0 = V(y_0)$ is the minimum value of the quasi-potential on the boundary. By Basic Result \ref{where}, if the system has a global potential $U(x)$, then $V_0/2 = U(y_0)-U(x_0)$, which represents the potential difference between the minimum point $y_0$ of the potential on the boundary and the attractor $x_0$. These facts can be seen from our mathematical derivations in \emph{Supplementary Information}.
\end{result}

It is a well-known result that the time needed for a Markov chain to make a state transition follows the exponential distribution. The Freidlin-Wentzell theory tells us that the escape time from each attraction basin approximately follows an exponential distribution. This shows that if we combine each attraction basin into a state, then the stochastic system with multiple attractors can be approximated by a Markov chain with multiple states at the time scale of $\exp(1/\epsilon)$. When the noise level $\epsilon$ is small, $\exp(1/\epsilon)$ becomes very large. This shows that the approximate Markov chain reflects the large-time-scale dynamic behavior of the stochastic system.

According to the above discussion, each stochastic model of a multistable system at the molecular level can be reduced to a Markov chain model at the cellular level. These two kinds of models at two different levels are essentially consistent and reflect the dynamic behavior of the system at two different time scales. For instance, the stochastic model \eqref{2d} proposed in this paper has two attraction basins when $a_0<a<a_1$, and thus can be approximated by a Markov chain model with two states, each corresponding to a phenotype. If the feedback architecture of the gene network becomes more complicated, then the stochastic system may possess three or more attraction basins and thus can be approximated by a Markov chain model with three or more states. For examples of Markov chain models with two, three, or four states, readers may refer to \cite{gaal2010exact, gupta2011stochastic, libby2011exclusion}.

\subsection*{Stochastic phenotype switching}
To survive in rapidly changing environments, a heterogenous bacterial population may allow individual cells to stochastically switch among multiple phenotypes, ensuring that some cells are always prepared for an unforeseen environmental fluctuation. This kind of phenotype switching is stochastic and temporary: An individual cell may switch to an alternative state at a random time and switch back again after some random time. Even without a significant change in environmental conditions, stochastic phenotype switching still exists. Stochastic phenotype switching has been observed in a wide range of bacterial species. As an example, upon encountering nutrient limitation, a minority of \emph{Bacillus subtilis} cells transiently enter the competent state with the capability for DNA uptake from the environment before returning to vegetative growth \cite{suel2006excitable}.

\begin{figure}[!htb]
\begin{center}
\centerline{\includegraphics[width=0.8\textwidth]{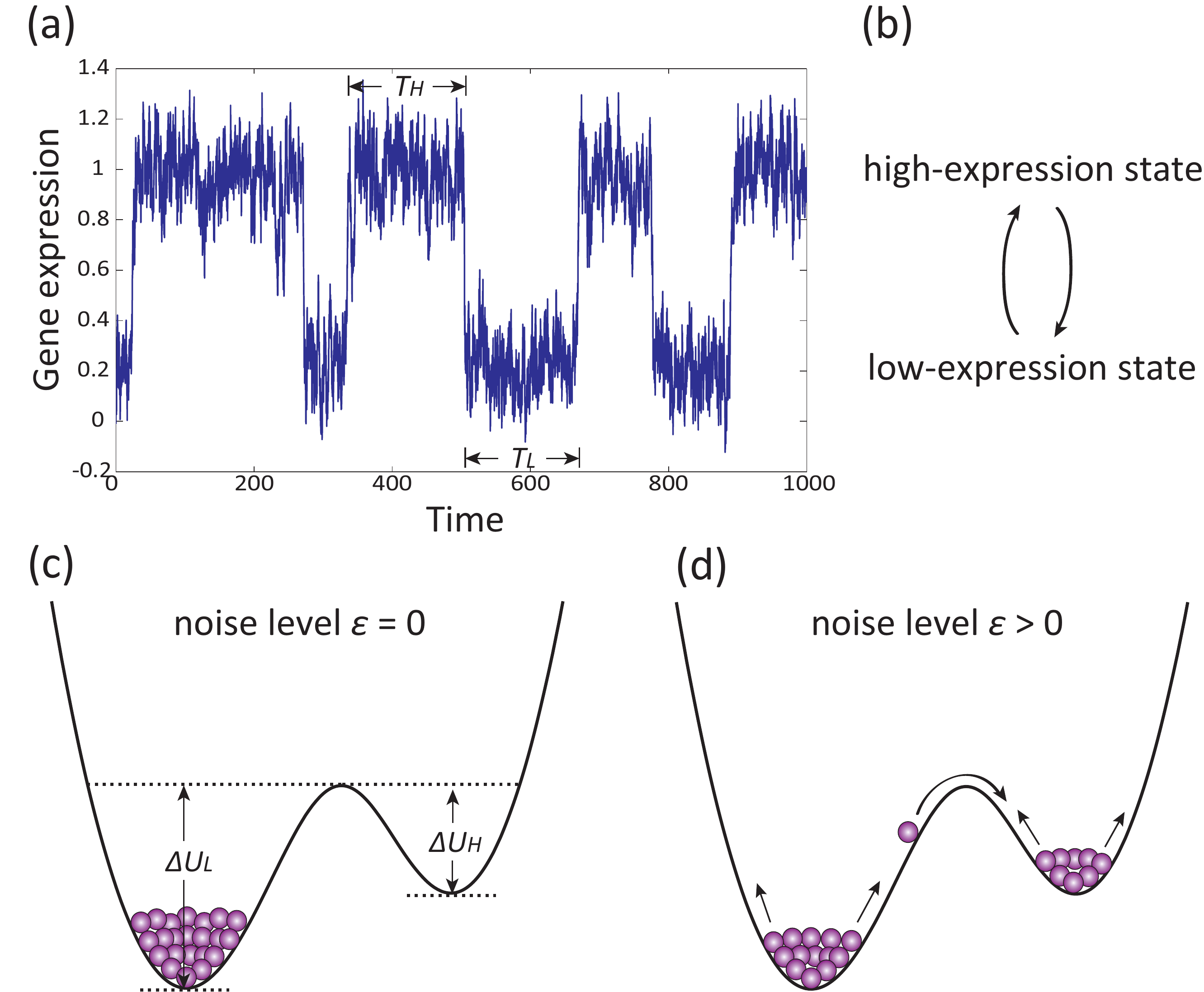}}
\caption{\textbf{Stochastic phenotype switching driven by stochastic forces.}
\textbf{a.} The simulation data of the time course of the level of protein X in a single cell. The cell switches between the low- and high-expression states at certain random times.
\textbf{b.} The simplified dynamics of the stochastic system \eqref{2d} as a two-state Markov chain.
\textbf{c.} The steady-state behavior of the system when the noise level $\epsilon$ is zero. The curve represents the one-dimensional effective potential $U(a,x,G(x))$. If the low-expressing cells are sorted out at a particular time, then all cells will stay in the low-expression state forever and phenotype switching is impossible.
\textbf{d.} The steady-state behavior of the system when the noise level $\epsilon$ is positive. The accumulation of stochastic forces will drive individual cells to surmount the potential barrier and make a state transition.}\label{switch}
\end{center}
\end{figure}

To better understand the principle behind stochastic phenotype switching, we simulate the time course of the level of protein X in an individual cell based on the stochastic system \eqref{2d} (Fig. \ref{switch}(a)). The simulation result shows that the cell switches between the low- and high-expression states at certain random times. The Freidlin-Wentzell theory shows that the escape times $T_L$ and $T_H$ from the low- and high-expression states approximately follow exponential distributions. The mean escape times, $\langle T_L\rangle$ and $\langle T_H\rangle$, have the following form (see \emph{Supplementary Information} and Basic Result \ref{when}):
\begin{equation}\label{escape}
\begin{split}
\langle T_L\rangle &\doteq \frac{2\pi}{\sqrt{\kappa_L\kappa_M}}\exp\left(\frac{1}{\epsilon}\Delta U_L\right),\\
\langle T_H\rangle &\doteq \frac{2\pi}{\sqrt{\kappa_H\kappa_M}}\exp\left(\frac{1}{\epsilon}\Delta U_H\right)
\end{split}
\end{equation}
where $\Delta U_L$ and $\Delta U_H$ are the potential differences between the saddle and the two attractors (Fig. \ref{switch}(c)) and $\kappa_L$, $\kappa_M$, and $\kappa_H$ are the curvatures of the one-dimensional effective potential $U(a,x,G(x))$ at $x_L$, $x_M$, and $x_H$, respectively. At the time scale of $\exp(1/\epsilon)$, the dynamic behavior of the stochastic system \eqref{2d} can be approximated by a Markov chain model with two states (Fig. \ref{switch}(b)).

According to Equation \eqref{escape}, the mean escape time is an exponential function of the potential barrier. The higher the potential barrier, the longer time is needed for a cell to make a state transition. Equation \eqref{escape} also shows that the mean escape times have the time scale of $\exp(1/\epsilon)$. This suggests that stochastic phenotype switching is a large-time-scale dynamic behavior of the system. When the noise level $\epsilon$ is small, the escape time may be longer than the time of cell division. For a fraction of cells, stochastic phenotype switching may not occur in a single cell cycle, and thus they can pass their phenotypic state to the next generation.

If there were no stochastic effects, the phenotype of each individual cell would never change. Specifically, if the low-expressing cells are sorted out at a particular time, then all cells will stay in the low-expression state forever and phenotype switching is impossible (Fig. 3(c)). In the presence of stochastic effects, however, the accumulation of stochastic forces will drive individual cells to surmount the potential barrier and make a state transition (Fig. \ref{switch}(d)). This suggests that stochasticity in gene expression is the driving force for stochastic phenotype switching.

\subsection*{Importance of the critical state}
We have seen that at certain ranges of the inducer concentration, the two attractors of the stochastic system \eqref{2d} are separated by a boundary, forming two attraction basins. The saddle of the system lies exactly on the boundary of the two attraction basins (Fig. \ref{SDEmodel}(b)) and thus characterizes a critical state between the two steady states of gene expression. This saddle is not targeted in the previous work since it is rarely observed in experiments and cannot be estimated by simple statistical analysis of gene expression data. However, the identification of the critical state has drawn increasing attention in recent years due to the following three reasons.

First, the saddle represents a critical level of gene expression. Recent studies on complex diseases show that any disease progression can be divided into a normal state, a pre-disease state, and a disease state \cite{kim2007hidden, kellershohn2001prion}, similar to the low-expression state, the critical state, and the high-expression state described in this paper. Once the expression levels of the disease-related genes in a person is close to the saddle, we have good reasons to believe that this person is in a pre-disease state and is at high risk of disease progression. This suggests that the identification of the critical state is closely related to the early diagnosis of complex diseases.

Second, the saddle is the most important point on the boundary of two adjacent attraction basins. According to the Freidlin-Wentzell theory, when the system escapes from one attraction basin to another, it must cross the boundary around a specific point where the quasi-potential attains its minimum. To see this, we simulate the time course of the levels of proteins X and Y in an individual cell based on the stochastic system \eqref{2d} (Fig. \ref{SDEmodel}(b)). The simulation result shows that the protein levels of the cell stay around the attractors at most times and cross the boundary of two attraction basins around the saddle, which is exactly the minimum point of the potential $U(a,x,y)$ on the boundary.

Third, the saddle characterizes a critical state of the transition between multiple attraction
basins. In order to accomplish stochastic phenotype switching, the system needs first to climb up the potential from one attractor to the saddle, and then to fall down the potential from the saddle to another attractor. Before reaching the saddle, the accumulation of stochastic forces will drive the system to climb up the potential against the potential gradient. This process in general will take rather a long time. Once the system crosses the saddle, it will reach another attractor along the potential gradient in a very short time. This shows that the dynamic features of a multistable system before and after reaching the saddle are totally different. To be more precise, let $T_u$ denote the time needed for the system to climb up the potential and let $T_d$ denote the time needed for the system to fall down the potential. The Freidlin-Wentzell theory shows that the ratio of $T_u$ to $T_d$ has the time scale of $\exp(1/\epsilon)$. This suggests that the process of climbing up the potential is much longer than that of falling down the potential, which is consistent with the old saying: Diseases come on horseback, but go away on foot.

\subsection*{Identification of the critical state}
The critical state of a multistable system is important in many ways. Thus it is natural to ask whether we can identify the critical state in an effective way based on the noisy data of gene expression. Recently, Chen et al. \cite{liu2012identifying} have developed a method of identifying the leading network in complex diseases by evaluating a kind of network entropy, and Gore et al. \cite{dai2012generic} have used generic statistical indicators to provide early warning signals for catastrophic collapse in the budding yeast population. Inspired by their ideas, in this section, we shall develop an effective method to identify the critical state of a multistable system using the time-course data of gene expression. Moreover, we shall validate the effectiveness of our method through both theoretical derivations and numerical simulation.

\begin{figure}[!htb]
\begin{center}
\centerline{\includegraphics[width=0.9\textwidth]{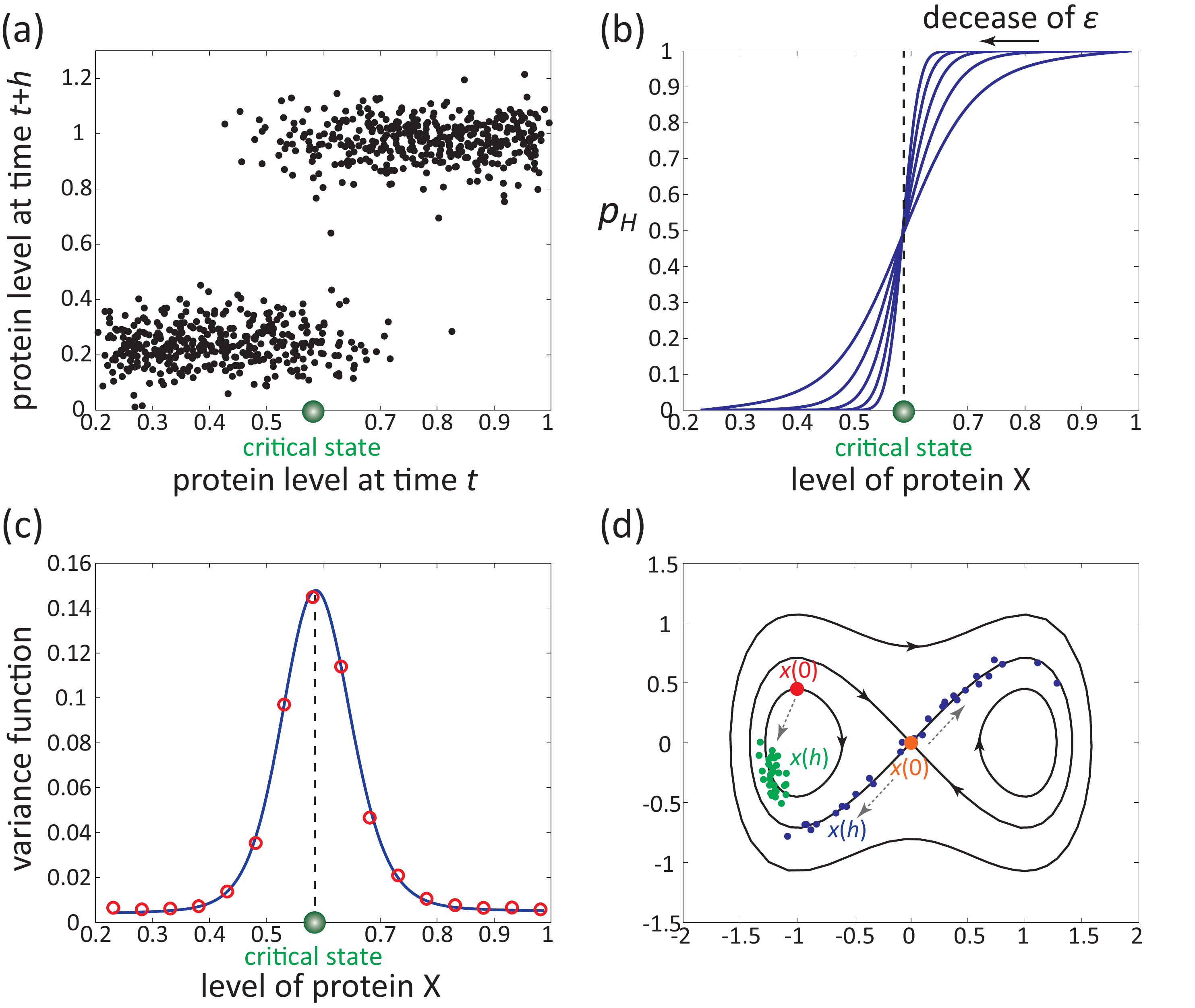}}
\caption{\textbf{Identification of the critical state.}
\textbf{a.} The simulation result of the level of protein X at two times, $t$ and $t+h$, for an ensemble of 760 samples. The samples at time $t+h$ are rather concentrated if the protein level is around $x_L$ or $x_H$ at time $t$, whereas those are rather scattered if the protein level is around $x_M$ at time $t$.
\textbf{b.} The graph of the function $p_H(x)$. The function $p_H(x)$ experiences a critical transition around the critical state $x_M$. With the decrease of the noise level $\epsilon$, the slope of the function $p_H(x)$ at $x_M$ tends to infinity.
\textbf{c.} The variance function $D(x)$. The variance $D(x)$ at $x$ is defined as the variance of the level of protein X at time $t+h$ conditioned on the information that the protein level equals $x$ at time $t$. The blue line represents the theoretical curve of the variance function and each red circle represents the simulation result based on the simulated dynamics of 300 virtual cells. The variance function changes slowly around the stable fixed points, $x_L$ and $x_H$, and experiences a drastic change around the critical state $x_M$.
\textbf{d.} A multistable stochastic system with periodic orbits. The lines with arrows give the phase portrait the system. The origin (0,0) (orange dot) is the unique critical point (saddle) of the system. The system has many periodic orbits inside and outside the two homoclinic orbits. We simulate the dynamics of the system at time $h = 3.5$ for an ensemble of 60 samples under two different initial values (red and orange dots). The green dots represent the samples at time $h$ when the system starts from the point (-1,0.45) (red dot) on a periodic orbit and the blue dots represent the samples at time $h$ when the system starts from the critical point (orange dot).
}\label{critical}
\end{center}
\end{figure}

We have seen that at certain ranges of the inducer concentration, the pivotal gene, gene \emph{X}, has two steady-state expression levels, $x_L$ and $x_H$, and a critical expression level $x_M$. We assume that we have measured the level of protein X in each individual cell within a bacterial population at two times, $t$ and $t+h$, where the interval $h$ of two successive measurements is chosen to have the time scale of $1/\epsilon$, which is much shorter than the time scale $\exp(1/\epsilon)$ of stochastic phenotype switching. Intuitively, if the protein level in an individual cell is around $x_L$ or $x_H$ at time $t$, then the protein level at time $t+h$ should be also around $x_L$ or $x_H$ since the interval $h$ is much shorter than the time needed for stochastic phenotype switching (Fig. \ref{critical}(a)). However, if the protein level in an individual cell is around $x_M$ at time $t$, then the protein level at time $t+h$ will become rather scattered since the critical state is rather unstable (Fig. \ref{critical}(a)).

Let $x(t)$ denote the level of protein X at time $t$, whose value can differ significantly between two individual cells. The above discussion illuminates us to define the variance $D(x)$ at $x$ as the variance of $x(t+h)$ conditioned on the information of $x(t)=x$. More precisely, we define the variance $D(x)$ at $x$ as
\begin{equation}
D(x) = \textrm{Variance}(x(t+h)|x(t) = x),
\end{equation}
where $\textrm{Variance}(Z) = \langle(Z-\langle Z\rangle)^2\rangle$ is the variance of the random variable $Z$. According to the above discussion, the variance $D(x)$ around $x_L$ or $x_H$ should be small since the distribution of $x(t+h)$ will be rather concentrated if $x(t)$ is around $x_L$ or $x_H$, whereas the variance $D(x)$ around $x_M$ should be large since the distribution of $x(t+h)$ will be rather scattered if $x(t)$ is around $x_M$. This suggests that we may detect the critical state by seeking the maximum point of the variance function $D(x)$.

We next validate the above intuitive discussion from the theoretical point of view. In fact, the theoretical expression of the variance function $D(x)$ has the following form (see \emph{Supplementary Information}):
\begin{equation}\label{variance}
D(x) \doteq (x_H-x_L)^2p_L(x)p_H(x) + \epsilon\left(\frac{p_L(x)}{\kappa_L}+\frac{p_H(x)}{\kappa_H}\right),
\end{equation}
where $p_L(x)$ and $p_H(x)$ are two functions (see \emph{Supplementary Information} for specific expressions) satisfying $p_L(x)+p_H(x) = 1$. We denote the maximum point of the variance function $D(x)$ by $x_{max}$. By Equation \eqref{variance}, we can easily see that $x_{max}$ satisfies
\begin{equation}\label{peakD}
p_H(x_{max}) = \frac{1}{2}+ \frac{(\kappa_L-\kappa_H)\epsilon}{2(x_H-x_L)^2\kappa_L\kappa_H}.
\end{equation}
In order to find the location of the maximum point $x_{max}$, we depict the graph of the function $p_H(x)$ in Fig. \ref{critical}(b), from which we see that the function $p_H(x)$ is sigmoidal with a critical transition around $x_M$. With the decrease of the noise level $\epsilon$, the slope of the function $p_H(x)$ at $x_M$ tends to infinity. This fact, together with Equation \eqref{peakD}, indicates that when the noise level $\epsilon$ is small, the maximum point $x_{max}$ of the variance function $D(x)$ is very close to $x_M$. This suggests that the variance function $D(x)$ may provide a clear signal for the position of the critical state.

A natural and important question is that whether we can estimate the variance function $D(x)$ from the noisy data of gene expression. The answer is of course affirmative. Recent experimental techniques such as fluorescent labeling and microfluidic devices allow us to measure the expression level of the pivotal gene in each individual cell within a bacterial population at a series of times $t_1,\cdots,t_m$ with interval $h$. We assume that the experimental data of the $n$-th cell have the form of $x(n,t_1),\cdots,x(n,t_m)$, where $x(n,t_i)$ represents the expression level of the $n$-th cell at time $t_i$. For each time $t_i$, if $x(n,t_i)$ is in a given small neighborhood of $x$, we pick out its next datum $x(n,t_{i+1})$, whereas if $x(n,t_i)$ is not in the given small neighborhood of $x$, we throw away its next datum. By evaluating the sample variance of those data which are picked out, we can obtain a good estimation of the variance $D(x)$ at $x$.

To validate the effectiveness of our method, we simulate the time course of the level of protein X in 4800 virtual cells based on the stochastic model \eqref{2d} and estimate the variance $D(x)$ at several discrete levels of $x$ (red circles in Fig. \ref{critical}(c)), from which we see that the simulation result coincides perfectly with the theoretical result (blue line in Fig. \ref{critical}(c)). Moreover, we see that the variance function $D(x)$ changes drastically around $x_M$, whereas there is no significant change in $D(x)$ around $x_L$ and $x_H$ (Fig. \ref{critical}(c)). This suggests that our method is effective in detecting the critical state of a multistable system using the noisy data of gene expression, even if no detailed mathematical model is available.

Before leaving this section, we would like to point out that our method of identifying the critical state is not only effective in simple multistable systems with stable fixed points, but also effective in complicated multistable systems with periodic orbits or stable limit cycles, in which case the time-course data of the system will oscillate. To see this, we consider the following two-dimensional system of SDEs:
\begin{equation}\label{cycle}\left\{
\begin{split}
\dot{x} &= y + \sqrt{2\epsilon}\xi_x, \\
\dot{y} &= -x^3+x + \sqrt{2\eta}\xi_y,
\end{split}\right.
\end{equation}
where $\epsilon$ and $\eta$ are two noise levels and $\xi_x$ and $\xi_y$ are two independent standard white noises. If $\epsilon = \eta = 0$, then the stochastic system is reduced to a deterministic system whose phase portrait is illustrated in Fig. \ref{critical}(d). The origin (0,0) (orange dot) is the only saddle and thus is the only critical point of the system. Interestingly, the stable and unstable manifolds at the critical point match up exactly, forming two homoclinic orbits. Moreover, the system has many periodic orbits inside and outside the two homoclinic orbits.

In order to see whether our method can be applied to identify the critical point of the stochastic system \eqref{cycle}, we simulate the dynamics of the system at time $h = 3.5$ for an ensemble of 60 samples under two different initial values (red and orange dots) (Fig. \ref{critical}(d)). The simulation result shows that the samples at time $h$ (green dots) are rather concentrated if the system starts from the point (-1,0.45) (red dot) on a periodic orbit, whereas the samples at time $h$ (blue dots) become rather scattered if the system starts from the critical point (orange dot). This suggests that the variance at the critical point is much larger than the variances at the points lying on the periodic orbits. Accordingly, we see that our method of identifying the critical state by seeking the maximum point of the variance function may be applied to various complicated multistable systems, not restricted to simple multistable systems with stable fixed points, such as the stochastic system \eqref{2d}.

\section*{Discussion}

\subsection*{Comparison with the previous work}
Multistability in biological systems has been widely studied in recent two decades and it has become an important recurring theme in cell signaling. In this paper, we study a class of multistable biological systems, i.e., heterogeneous bacterial populations with stochastic phenotype switching and also provide a data-driven method to identify the critical state of multistable bacterial systems. In recent literature, an influential paper about multistability is the paper of Angeli, Ferrell, and Sontag \cite{angeli2004detection}, who studied a large class of biological systems with positive feedbacks and also provided a method of detecting bistability, bifurcations, and hysteresis in these systems. The themes of their paper is closely related to ours. Thus we find it necessary to discuss the similarities and differences between their paper and ours.

First, both the paper of Angeli et al. and ours studied multistability in biological systems with a positive-feedback network. Recent studies show that multistability always arises in biological systems that contain a positive-feedback loop and it has been proved that the existence of at least one positive-feedback loop is a necessary condition for the existence of multiple steady states \cite{gouze1998positive, snoussi1998necessary, cinquin2002positive}. In our paper, we show that phenotypic heterogeneity in bacterial populations often results from the underlying double-positive-feedback gene networks, which is consistent with the above theoretical results.

Second, the paper of Angeli et al. studied the deterministic nonlinear dynamics of multistable biological systems and our paper studies the stochastic nonlinear dynamics of heterogenous bacterial populations. In fact, deterministic models may lead to phenotypic heterogeneity, but they cannot explain the widely observed phenomena of stochastic phenotype switching and bet-hedging. In our paper, we use our stochastic model to provide a clear description of stochastic phenotype switching and bet-hedging within heterogenous bacterial populations and study the role that stochastic effects play in generating these important experimental phenomena.

Third, in the paper of Angeli et al., the relationship between the deterministic model of multistable systems and the widely used Markov chain model of population evolution is not clear. In our paper, we use the Freidlin-Wentzell theory to show that our stochastic model at the molecular level can be approximated by a Markov chain model at the cellular level, which reflects the large-time-scale dynamics of multistable bacterial systems, when the noise level is small. It turns out that only by considering the stochastic dynamics of multistable bacterial systems can we unify the models at the two different levels.

Fourth, the paper of Angeli et al. provided a possible method to detect multistability and bifurcations in a class of multistable systems satisfying the so-called ``monotonicity" and ``open-loop steady-state response" assumptions. Their method strongly depends on the properties of the system when the feedback is blocked. In order to detect multistability and bifurcations, the response data of the open-loop, feedback-blocked system to input stimuli must be obtained. This requirement is obvious too strong for realistic biological systems. In our paper, however, we provide an effective data-driven method to identify the critical state of multistable biological systems. In our method, only the time-course data of gene expression in individual cells are needed, even if no detailed mathematical model is available.

\subsection*{Rationality of our stochastic model}
In our stochastic model, we have used white noises with two different noise levels to describe the stochastic fluctuations in proteins X and Y, respectively. In this subsection, we shall explain the rationality of this assumption. The content of this section is somewhat technical. Readers who are unfamiliar with the knowledge of stochastic processes can skip this part.

In fact, the most precise model of the gene network in living cells is the chemical master equation (CME) \cite{gillespie1992rigorous}. Mathematically, the CME is the equation satisfied by the probability distribution of a Markov jump process which describes the copy number fluctuations of all participating macromolecules in the gene network. However, the dimension of the CME model is often so high that the theoretical expressions of many important quantities related to the system cannot be explicitly calculated. This makes the CME model difficult to be directly applied to practical problems with experimental data and observations.

In order to simplify the CME model, Kurtz \cite{kurtz1970solutions, kurtz1971limit, kurtz1972relationship} proved in his pioneering work that every CME model can be approximated reasonably well by the so-called chemical Langevin equation when the volume of the system is large. Mathematically, the chemical Langevin equation \cite{gillespie2000chemical} is the equation satisfied by the probability distribution of an SDE model which describes the concentration fluctuations of all participating macromolecules in the gene network. To be more precise, we assume that there are $n$ macromolecules involved in the gene network whose concentrations are denoted by $x_1,x_2,\cdots,x_n$. The general form of an SDE model which governs $x_1,x_2,\cdots,x_n$ is given by
\begin{equation}\label{diffusion}
\dot{x}_i = b_i(x) + \sqrt{2\epsilon_i}\sum_{j=1}^n\sigma_{ij}(x)\xi_j,\;\;\;i=1,2,\cdots,n,
\end{equation}
where $x=(x_1,x_2,\cdots,x_n)$, $b(x)=(b_1(x),b_2(x),\cdots,b_n(x))$ is the drift coefficient of the SDE, $\sigma(x)=(\sigma_{ij}(x))_{n\times n}$ is the diffusion coefficient of the SDE, and $\xi_1,\xi_2,\cdots,\xi_n$ are $n$ independent standard white noises. Mathematically, it can be proved that when the matrix $\sigma(x)$ is bounded from both below and above, almost all the major properties of the SDE model related to the Freidlin-Wentzell theory will change little if $\sigma(x)$ is regarded as the identity matrix \cite{freidlin2012random, olivieri2005large}, in which case the SDE model \eqref{diffusion} can be simplified as
\begin{equation}\label{simplified}
\dot{x}_i = b_i(x) + \sqrt{2\epsilon_i}\xi_i,\;\;\;i=1,2,\cdots,n.
\end{equation}
In fact, the SDE has been applied to model the concentration fluctuations of macromolecules in gene networks in some previous studies \cite{suel2006excitable}. Some reviews on this topic can be found in \cite{rao2002control, wilkinson2009stochastic, meister2014modeling}.

In the core double-positive-feedback gene network depicted in Fig. \ref{network}(e), there are only two participating macromolecules, proteins X and Y, if the inducer concentration is regarded as a parameter. In this case, the simplified SDE model \eqref{simplified} is exactly our two-dimensional stochastic model \eqref{2d}. To make biologists, especially experimental biologists, understand the main results of this paper, we would like to use white noises to describe stochasticity in gene expression and use the simplified SDE \eqref{simplified} to model multistable bacterial systems. Under this simplification, our stochastic model \eqref{2d} can be simply understood as the random perturbation of the traditional deterministic model.

\subsection*{Strengths and deficiencies of the SDE model}
Just as George Box's famous saying said: all models are wrong, but some are useful. Our model is no exception. In this paper, we use the SDE to model the concentration fluctuations of the participating macromolecules in the gene network of heterogenous bacterial populations. However, the SDE model proposed in this paper has some deficiencies. In fact, in living bacterial systems, some macromolecules, such as mRNA and protein molecules, may exist at very low copy numbers \cite{xie2008single}, in which case the concept of concentration makes no sense. Thus the SDE model cannot provide a good approximation of the CME model when the copy numbers of the participating macromolecules are very small.

However, compared with the CME model, the SDE model proposed in this paper has many strengths. First, the dimension of the CME model is often too high to be directly applied to practical problems with experimental data and observations. If the gene network of the bacterial system contains $n$ macromolecules and the maximal possible copy numbers of these $n$ macromolecules are $N_1,N_2,\cdots, N_n$. Then the dimension of the CME model will be $N_1N_2\cdots N_n$. However, the SDE model compresses the dimension of the system to a large extent. If the gene network of the bacterial system contains $n$ macromolecules, then the dimension of the SDE model is only $n$. This makes the system easy to be analyzed and makes many important quantities related to the system easy to be explicitly calculated.

Second, the mathematical theory of the SDE model is well developed, whereas that of the CME model is poorly developed due to its high complexity. In this paper, we use the Freidlin-Wentzell theory to explain stochastic phenotype switching and bet-hedging within isogenic bacterial populations and the consistence between our stochastic model and the traditional Markov chain models of population evolution. We believe that an analogue of the Freidlin-Wentzell theory must exist in the CME model. However, such a theory for the CME model is not well developed up till now.

Third, the CME model is so abstract that it is not easy for biologists to understand and follow. In fact, without some deep mathematical knowledge, it is very difficult to understand the relationship between the CME model and the traditional deterministic model. However, the SDE model can be simply viewed as the random perturbation of the deterministic model. This makes the SDE model easy to be understood for both theoretical and experimental biologists.

Due to these reasons, we choose to use the SDE to model heterogenous bacterial populations, instead of the CME. We believe that the SDE, or equivalently, the chemical Langevin equation, which does not lose much information of the complicated CME, is a useful tool in the modeling of gene regulatory networks with inherent noises and in the analysis of the stochastic nonlinear dynamics of biological systems.

\subsection*{Potential of our stochastic model}
Given the small size of a cell and the small copy numbers of participating macromolecules, cellular processes during gene expression are inherently stochastic. In this paper, we establish a unified nonlinear stochastic model of multistable bacterial systems at the molecular level based on a core double-positive-feedback gene network (Fig. \ref{network}(e)) and provide a clear description of phenotypic heterogeneity, stochastic phenotype switching, and bet-hedging within isogenic bacterial populations. Although we have used the expression levels of two stress-related genes to establish our model, we point out here that the variables in our model are not necessarily the expression levels of the stress-related genes, but can include more comprehensive indicators measured by flow cytometry and other techniques, such as FSC (roughly proportional to cell size), SSC (roughly proportional to cell granularity and complexity), and the ATP concentration (positively correlated with total cell energy).

In our recent study about antibiotic resistance of \emph{Escherichia coli} (unpublished work), we found that the one-dimensional expression data of the hydrolase gene only lead to a monomodal distribution, but the multidimensional expression data of a group of stress-related genes lead to an apparent multimodal distribution. This phenomenon is described in this paper. Our simulation result shows that although the expression data of a group of genes are distributed within multiple attraction basins, their marginal distribution may overlap to a large extent so that we may not be able to observe a multimodal distribution if we only focus on the expression data of a single gene. In our future work, we shall further apply the general framework discussed in this paper to the specific problem of antibiotic resistance of \emph{Escherichia coli}.

Biological systems with multistability are ubiquitous in nature. Some fundamental cellular processes, such as decision-making processes in cell cycle progression \cite{yao2008bistable}, cell fate determination \cite{ferrell1998biochemical, xiong2003positive, corson2012geometry, yan2013single}, and apoptosis \cite{eissing2004bistability, spencer2009non}, display multistable features. In addition, multistability is also involved in disease progression, which can be divided into a normal state, a pre-disease state, and a disease state \cite{kellershohn2001prion, kim2007hidden}. Although various multistable biological systems have different feedback regulatory networks, the mathematical structures behind them are quite similar. We hope that the stochastic approach discussed in this paper can give enlightenment to the understanding of biological systems with multistability and to the analysis of the related new phenomena and new questions.

\section*{Methods}
The parameters used to draw the figures in the main text are chosen as $\alpha = 1$, $\beta = 2$, $\gamma = 0.8$, $K = 0.05$, $\delta = 0$, and $n = 6$. The inducer concentration $a$ and the two noise levels, $\epsilon$ and $\eta$, are chosen appropriately when drawing different figures.

\section*{Acknowledgements}
The authors gratefully acknowledge Prof. Michael S. Waterman at the University of Southern California and Prof. Michael Q. Zhang at Tsinghua University for stimulating discussions. This work is supported by NSFC 11271029 and NSFC 11171024.

\setlength{\bibsep}{5pt}
\small\bibliographystyle{pnas2009}
\bibliography{phenotype}

\begin{thebibliography}{10}

\bibitem{kussell2005phenotypic}
Kussell E, Leibler S (2005) Phenotypic diversity, population growth, and
  information in fluctuating environments.
\newblock \emph{Science} 309:2075--2078.

\bibitem{smits2006phenotypic}
Smits WK, Kuipers OP, Veening JW (2006) Phenotypic variation in bacteria: the
  role of feedback regulation.
\newblock \emph{Nat Rev Microbiol} 4:259--271.

\bibitem{dubnau2006bistability}
Dubnau D, Losick R (2006) Bistability in bacteria.
\newblock \emph{Mol Microbiol} 61:564--572.

\bibitem{avery2006microbial}
Avery SV (2006) Microbial cell individuality and the underlying sources of
  heterogeneity.
\newblock \emph{Nat Rev Microbiol} 4:577--587.

\bibitem{dhar2007microbial}
Dhar N, McKinney JD (2007) Microbial phenotypic heterogeneity and antibiotic
  tolerance.
\newblock \emph{Curr Opin Microbiol} 10:30--38.

\bibitem{lu2007phenotypic}
Lu T, Shen T, Bennett MR, Wolynes PG, Hasty J (2007) Phenotypic variability of
  growing cellular populations.
\newblock \emph{Proc Natl Acad Sci U S A} 104:18982--18987.

\bibitem{veening2008bistability}
Veening JW, Smits WK, Kuipers OP (2008) Bistability, epigenetics, and
  bet-hedging in bacteria.
\newblock \emph{Annu Rev Microbiol} 62:193--210.

\bibitem{fraser2009chance}
Fraser D, K{\ae}rn M (2009) A chance at survival: gene expression noise and
  phenotypic diversification strategies.
\newblock \emph{Mol Microbiol} 71:1333--1340.

\bibitem{jablonka2009transgenerational}
Jablonka E, Raz G (2009) Transgenerational epigenetic inheritance: prevalence,
  mechanisms, and implications for the study of heredity and evolution.
\newblock \emph{Q Rev Biol} 84:131--176.

\bibitem{snijder2011origins}
Snijder B, Pelkmans L (2011) Origins of regulated cell-to-cell variability.
\newblock \emph{Nat Rev Mol Cell Biol} 12:119--125.

\bibitem{mao2014slow}
Mao J, Blanchard AE, Lu T (2014) Slow and steady wins the race: A bacterial
  exploitative competition strategy in fluctuating environments.
\newblock \emph{ACS Synth Biol} .

\bibitem{rulands2014specialization}
Rulands S, Jahn D, Frey E (2014) Specialization and bet hedging in
  heterogeneous populations.
\newblock \emph{Phys Rev Lett} 113:108102.

\bibitem{acar2008stochastic}
Acar M, Mettetal JT, van Oudenaarden A (2008) Stochastic switching as a
  survival strategy in fluctuating environments.
\newblock \emph{Nat Genet} 40:471--475.

\bibitem{salathe2009evolution}
Salath{\'e} M, Van~Cleve J, Feldman MW (2009) Evolution of stochastic switching
  rates in asymmetric fitness landscapes.
\newblock \emph{Genetics} 182:1159--1164.

\bibitem{leisner2008stochastic}
Leisner M, Stingl K, Frey E, Maier B (2008) Stochastic switching to competence.
\newblock \emph{Curr Opin Microbiol} 11:553--559.

\bibitem{gaal2010exact}
Ga{\'a}l B, Pitchford JW, Wood AJ (2010) Exact results for the evolution of
  stochastic switching in variable asymmetric environments.
\newblock \emph{Genetics} 184:1113--1119.

\bibitem{libby2011exclusion}
Libby E, Rainey PB (2011) Exclusion rules, bottlenecks and the evolution of
  stochastic phenotype switching.
\newblock \emph{P Roy Soc B - Biol Sci} 278:3574--3583.

\bibitem{rainey2011evolutionary}
Rainey PB, et~al. (2011) The evolutionary emergence of stochastic phenotype
  switching in bacteria.
\newblock \emph{Microb Cell Fact} 10:S14.

\bibitem{ozbudak2004multistability}
Ozbudak EM, Thattai M, Lim HN, Shraiman BI, Van~Oudenaarden A (2004)
  Multistability in the lactose utilization network of escherichia coli.
\newblock \emph{Nature} 427:737--740.

\bibitem{suel2006excitable}
S{\"u}el GM, Garcia-Ojalvo J, Liberman LM, Elowitz MB (2006) An excitable gene
  regulatory circuit induces transient cellular differentiation.
\newblock \emph{Nature} 440:545--550.

\bibitem{tsang2006exciting}
Tsang J, Van~Oudenaarden A (2006) Exciting fluctuations: monitoring competence
  induction dynamics at the single-cell level.
\newblock \emph{Mol Syst Biol} 2.

\bibitem{schultz2007molecular}
Schultz D, Jacob EB, Onuchic JN, Wolynes PG (2007) Molecular level stochastic
  model for competence cycles in bacillus subtilis.
\newblock \emph{Proc Natl Acad Sci U S A} 104:17582--17587.

\bibitem{sonenshein2002bacillus}
Sonenshein AL, Hoch JA, Losick R (2002) \emph{Bacillus subtilis and its closest
  relatives: from genes to cells} (Asm Press Washington DC:).

\bibitem{errington2003regulation}
Errington J, et~al. (2003) Regulation of endospore formation in bacillus
  subtilis.
\newblock \emph{Nat Rev Microbiol} 1:117--126.

\bibitem{morohashi2007model}
Morohashi M, et~al. (2007) Model-based definition of population heterogeneity
  and its effects on metabolism in sporulating bacillus subtilis.
\newblock \emph{J Biochem (Tokyo)} 142:183--191.

\bibitem{de2010heterochronic}
de~Jong IG, Veening JW, Kuipers OP (2010) Heterochronic phosphorelay gene
  expression as a source of heterogeneity in bacillus subtilis spore formation.
\newblock \emph{J Bacteriol} 192:2053--2067.

\bibitem{sureka2008positive}
Sureka K, et~al. (2008) Positive feedback and noise activate the stringent
  response regulator rel in mycobacteria.
\newblock \emph{PLoS One} 3:e1771.

\bibitem{gefen2009importance}
Gefen O, Balaban NQ (2009) The importance of being persistent: heterogeneity of
  bacterial populations under antibiotic stress.
\newblock \emph{FEMS Microbiol Rev} 33:704--717.

\bibitem{ghosh2011phenotypic}
Ghosh S, et~al. (2011) Phenotypic heterogeneity in mycobacterial stringent
  response.
\newblock \emph{BMC Syst Biol} 5:18.

\bibitem{veening2008bet}
Veening JW, et~al. (2008) Bet-hedging and epigenetic inheritance in bacterial
  cell development.
\newblock \emph{Proc Natl Acad Sci U S A} 105:4393--4398.

\bibitem{beaumont2009experimental}
Beaumont HJ, Gallie J, Kost C, Ferguson GC, Rainey PB (2009) Experimental
  evolution of bet hedging.
\newblock \emph{Nature} 462:90--93.

\bibitem{wolf2005diversity}
Wolf DM, Vazirani VV, Arkin AP (2005) Diversity in times of adversity:
  probabilistic strategies in microbial survival games.
\newblock \emph{J Theor Biol} 234:227--253.

\bibitem{gupta2011stochastic}
Gupta PB, et~al. (2011) Stochastic state transitions give rise to phenotypic
  equilibrium in populations of cancer cells.
\newblock \emph{Cell} 146:633--644.

\bibitem{zhou2013population}
Zhou D, Wu D, Li Z, Qian M, Zhang MQ (2013) Population dynamics of cancer cells
  with cell state conversions.
\newblock \emph{Quantitative Biology} :1--8.

\bibitem{karmakar2007positive}
Karmakar R, Bose I (2007) Positive feedback, stochasticity and genetic
  competence.
\newblock \emph{Phys Biol} 4:29--37.

\bibitem{mitrophanov2008positive}
Mitrophanov AY, Groisman EA (2008) Positive feedback in cellular control
  systems.
\newblock \emph{Bioessays} 30:542--555.

\bibitem{mantzaris2007single}
Mantzaris NV (2007) From single-cell genetic architecture to cell population
  dynamics: quantitatively decomposing the effects of different population
  heterogeneity sources for a genetic network with positive feedback
  architecture.
\newblock \emph{Biophys J} 92:4271--4288.

\bibitem{vellela2009stochastic}
Vellela M, Qian H (2009) Stochastic dynamics and non-equilibrium thermodynamics
  of a bistable chemical system: the schl{\"o}gl model revisited.
\newblock \emph{J R Soc Interface} 6:925--940.

\bibitem{qian2010chemical}
Qian H, Bishop LM (2010) The chemical master equation approach to
  nonequilibrium steady-state of open biochemical systems: Linear
  single-molecule enzyme kinetics and nonlinear biochemical reaction networks.
\newblock \emph{Int J Mol Sci} 11:3472--3500.

\bibitem{qian2011nonlinear}
Qian H (2011) Nonlinear stochastic dynamics of mesoscopic homogeneous
  biochemical reaction systems -- an analytical theory.
\newblock \emph{Nonlinearity} 24:R19.

\bibitem{ge2011non}
Ge H, Qian H (2011) Non-equilibrium phase transition in mesoscopic biochemical
  systems: from stochastic to nonlinear dynamics and beyond.
\newblock \emph{J R Soc Interface} 8:107--116.

\bibitem{qian2012cooperativity}
Qian H (2012) Cooperativity in cellular biochemical processes: noise-enhanced
  sensitivity, fluctuating enzyme, bistability with nonlinear feedback, and
  other mechanisms for sigmoidal responses.
\newblock \emph{Annu Rev Biophys} 41:179--204.

\bibitem{qian2012mesoscopic}
Qian H, Ge H (2012) Mesoscopic biochemical basis of isogenetic inheritance and
  canalization: Stochasticity, nonlinearity, and emergent landscape.
\newblock \emph{Mol Cellu Biomech} 9:1--30.

\bibitem{mcadams1997stochastic}
McAdams HH, Arkin A (1997) Stochastic mechanisms in gene expression.
\newblock \emph{Proc Natl Acad Sci U S A} 94:814--819.

\bibitem{elowitz2002stochastic}
Elowitz MB, Levine AJ, Siggia ED, Swain PS (2002) Stochastic gene expression in
  a single cell.
\newblock \emph{Science Signalling} 297:1183.

\bibitem{ozbudak2002regulation}
Ozbudak EM, Thattai M, Kurtser I, Grossman AD, van Oudenaarden A (2002)
  Regulation of noise in the expression of a single gene.
\newblock \emph{Nat Genet} 31:69--73.

\bibitem{paulsson2004summing}
Paulsson J (2004) Summing up the noise in gene networks.
\newblock \emph{Nature} 427:415--418.

\bibitem{kaern2005stochasticity}
K{\ae}rn M, Elston TC, Blake WJ, Collins JJ (2005) Stochasticity in gene
  expression: from theories to phenotypes.
\newblock \emph{Nat Rev Genet} 6:451--464.

\bibitem{raser2005noise}
Raser JM, O'Shea EK (2005) Noise in gene expression: origins, consequences, and
  control.
\newblock \emph{Science} 309:2010--2013.

\bibitem{cai2006stochastic}
Cai L, Friedman N, Xie XS (2006) Stochastic protein expression in individual
  cells at the single molecule level.
\newblock \emph{Nature} 440:358--362.

\bibitem{yu2006probing}
Yu J, Xiao J, Ren X, Lao K, Xie XS (2006) Probing gene expression in live
  cells, one protein molecule at a time.
\newblock \emph{Science} 311:1600--1603.

\bibitem{xie2008single}
Xie XS, Choi PJ, Li GW, Lee NK, Lia G (2008) Single-molecule approach to
  molecular biology in living bacterial cells.
\newblock \emph{Annu Rev Biophys} 37:417--444.

\bibitem{sanchez2013regulation}
Sanchez A, Choubey S, Kondev J (2013) Regulation of noise in gene expression.
\newblock \emph{Annu Rev Biophys} 42:469--491.

\bibitem{freidlin2012random}
Freidlin MI, Wentzell AD (2012) \emph{Random perturbations of dynamical
  systems}, volume 260 (Springer).

\bibitem{kim2007hidden}
Kim D, Rath O, Kolch W, Cho K (2007) A hidden oncogenic positive feedback loop
  caused by crosstalk between wnt and erk pathways.
\newblock \emph{Oncogene} 26:4571--4579.

\bibitem{kellershohn2001prion}
Kellershohn N, Laurent M (2001) Prion diseases: dynamics of the infection and
  properties of the bistable transition.
\newblock \emph{Biophys J} 81:2517--2529.

\bibitem{olivieri2005large}
Olivieri E, Vares ME (2005) \emph{Large deviations and metastability}.
\newblock 100 (Cambridge University Press).

\bibitem{liu2012identifying}
Liu R, et~al. (2012) Identifying critical transitions and their leading
  biomolecular networks in complex diseases.
\newblock \emph{Sci Rep - UK} 2:813.

\bibitem{dai2012generic}
Dai L, Vorselen D, Korolev KS, Gore J (2012) Generic indicators for loss of
  resilience before a tipping point leading to population collapse.
\newblock \emph{Science} 336:1175--1177.

\bibitem{angeli2004detection}
Angeli D, Ferrell JE, Sontag ED (2004) Detection of multistability,
  bifurcations, and hysteresis in a large class of biological positive-feedback
  systems.
\newblock \emph{Proc Natl Acad Sci U S A} 101:1822--1827.

\bibitem{gouze1998positive}
Gouz{\'e} JL (1998) Positive and negative circuits in dynamical systems.
\newblock \emph{J Biol Syst} 6:11--15.

\bibitem{snoussi1998necessary}
Snoussi EH (1998) Necessary conditions for multistationarity and stable
  periodicity.
\newblock \emph{J Biol Syst} 6:3--9.

\bibitem{cinquin2002positive}
Cinquin O, Demongeot J (2002) Positive and negative feedback: striking a
  balance between necessary antagonists.
\newblock \emph{J Theor Biol} 216:229--241.

\bibitem{gillespie1992rigorous}
Gillespie DT (1992) A rigorous derivation of the chemical master equation.
\newblock \emph{Physica A} 188:404--425.

\bibitem{kurtz1970solutions}
Kurtz TG (1970) Solutions of ordinary differential equations as limits of pure
  jump markov processes.
\newblock \emph{J Appl Probab} 7:49--58.

\bibitem{kurtz1971limit}
Kurtz TG (1971) Limit theorems for sequences of jump markov processes
  approximating ordinary differential processes.
\newblock \emph{J Appl Probab} 8:344--356.

\bibitem{kurtz1972relationship}
Kurtz TG (1972) The relationship between stochastic and deterministic models
  for chemical reactions.
\newblock \emph{J Chem Phys} 57:2976--2978.

\bibitem{gillespie2000chemical}
Gillespie DT (2000) The chemical langevin equation.
\newblock \emph{J Chem Phys} 113:297--306.

\bibitem{rao2002control}
Rao CV, Wolf DM, Arkin AP (2002) Control, exploitation and tolerance of
  intracellular noise.
\newblock \emph{Nature} 420:231--237.

\bibitem{wilkinson2009stochastic}
Wilkinson DJ (2009) Stochastic modelling for quantitative description of
  heterogeneous biological systems.
\newblock \emph{Nat Rev Genet} 10:122--133.

\bibitem{meister2014modeling}
Meister A, Du C, Li YH, Wong WH (2014) Modeling stochastic noise in gene
  regulatory systems.
\newblock \emph{Quantitative Biology} 2:1--29.

\bibitem{yao2008bistable}
Yao G, Lee TJ, Mori S, Nevins JR, You L (2008) A bistable rb--e2f switch
  underlies the restriction point.
\newblock \emph{Nat Cell Biol} 10:476--482.

\bibitem{ferrell1998biochemical}
Ferrell~Jr JE, Machleder EM (1998) The biochemical basis of an all-or-none cell
  fate switch in xenopus oocytes.
\newblock \emph{Science} 280:895--898.

\bibitem{xiong2003positive}
Xiong W, Ferrell JE (2003) A positive-feedback-based bistable ¡®memory
  module¡¯that governs a cell fate decision.
\newblock \emph{Nature} 426:460--465.

\bibitem{corson2012geometry}
Corson F, Siggia ED (2012) Geometry, epistasis, and developmental patterning.
\newblock \emph{Proc Natl Acad Sci U S A} 109:5568--5575.

\bibitem{yan2013single}
Yan L, et~al. (2013) Single-cell rna-seq profiling of human preimplantation
  embryos and embryonic stem cells.
\newblock \emph{Nat Struct Mol Biol} 20:1131--1139.

\bibitem{eissing2004bistability}
Eissing T, et~al. (2004) Bistability analyses of a caspase activation model for
  receptor-induced apoptosis.
\newblock \emph{J Biol Chem} 279:36892--36897.

\bibitem{spencer2009non}
Spencer SL, Gaudet S, Albeck JG, Burke JM, Sorger PK (2009) Non-genetic origins
  of cell-to-cell variability in trail-induced apoptosis.
\newblock \emph{Nature} 459:428--432.

\end{thebibliography}
\end{document}